\def\RELEASE{1}  %
\def\ANON{0}     %
\def\SQUEEZE{0}  %

\documentclass[nonacm,sigconf,10pt]{acmart}
\PassOptionsToPackage{hyphens}{url}\usepackage{hyperref}

\usepackage[noend]{algpseudocode}
\usepackage{algorithmicx}
\usepackage{booktabs}
\usepackage{caption}
\usepackage{subcaption}
\usepackage{comment}
\usepackage[inline]{enumitem}
\usepackage{graphicx}
\usepackage{listings}
\usepackage{multirow}
\usepackage{xcolor}
\usepackage{xspace}
\usepackage{microtype}
\usepackage{tikz}
\usepackage{gnuplot-lua-tikz}
\usepackage{array}
\usepackage{tabularx}
\usepackage{pifont}

\microtypecontext{spacing=nonfrench}

\hypersetup{
  pdflang={en},
  pdfdisplaydoctitle}

\definecolor[named]{OurPurple}{cmyk}{0.55,1,0,0.15}
\definecolor[named]{OurDarkBlue}{cmyk}{1,0.58,0,0.21}
\hypersetup{
  colorlinks,
  linkcolor=OurPurple,
  citecolor=OurPurple,
  urlcolor=OurDarkBlue,
  filecolor=OurDarkBlue}

\if \SQUEEZE 1
\setlist[itemize]{
  leftmargin=*,
  itemsep=2pt,
  topsep=2pt}

\fi

\def\Snospace~{\S{}}

\if \RELEASE 1
  \def\NOTES{0}
\else
  \def\NOTES{1}
\fi
\if \NOTES 1
  \newcommand{\XXX}[1]{{\color{red}{XXX {#1}}}}
  \newcommand{\antoine}[1]{{\color{teal}{[\textbf{AK:} {#1}]}}}
  \newcommand{\hj}[1]{{\color{violet}{[\textbf{HJ:} {#1}]}}}
  \newcommand{\lijl}[1]{{\color{orange}{[\textbf{JL:} {#1}]}}}
  \newcommand{\todo}[1]{{\color{blue}{TODO: {#1}}}}
\else
  \newcommand{\XXX}[1]{}
  \newcommand{\antoine}[1]{}
  \newcommand{\hj}[1]{}
  \newcommand{\lijl}[1]{}
  \newcommand{\todo}[1]{}
\fi

\if \ANON 1
  \newcommand{\sys}{SepaNet\xspace}
\else
  \newcommand{\sys}{SplitSim\xspace}
\fi

\begin{document}
\date{}

\if \ANON 1
  \title[\sys]{Towards Practical Large-Scale Full-System Simulation for Systems Research}
\else
  \title[\sys]{\sys: Large-Scale Simulations for Evaluating Network Systems Research}
\fi

\if \ANON 1
  \author{Anonymous Submission \#658 (\pageref{page:last} pages)}
\else
  \author{Hejing Li}
  \orcid{0000-0001-6930-2419}
  \affiliation{
    \institution{Max Planck Institute for Software Systems}
    \country{Germany}}
  \email{hejingli@mpi-sws.org}

  \author{Praneeth Balasubramanian}
  \affiliation{
    \institution{BITS Pilani}
    \country{India}}

  \author{Marvin Meiers}
  \affiliation{
    \institution{Saarland University}
    \country{Germany}}

  \author{Jialin Li}
  \orcid{0000-0003-3530-7662}
  \affiliation{
    \institution{National University of Singapore}
    \country{Singapore}}
  \email{lijl@comp.nus.edu.sg}

  \author{Antoine Kaufmann}
  \orcid{0000-0002-6355-2772}
  \affiliation{
    \institution{Max Planck Institute for Software Systems}
    \country{Germany}}
  \email{antoinek@mpi-sws.org}
\fi

\begin{abstract}
  When physical testbeds are out of reach for evaluating a networked
  system, we frequently turn to simulation.
  In today's datacenter networks, bottlenecks are rarely at the
  network protocol level, but instead in end-host software or hardware
  components, thus current protocol-level simulations are inadequate
  means of evaluation.
  End-to-end simulations covering these components on the other hand,
  simply cannot achieve the required scale with feasible simulation
  performance and computational resources.

  In this paper, we address this with \sys, a simulation
  framework for end-to-end evaluation for large-scale network
  and distributed systems.
  To this end, \sys builds on prior work on modular end-to-end
  simulations and combines this with key elements to achieve
  scalability.
  First, mixed fidelity simulations judiciously reduce detail in
  simulation of parts of the system where this can be tolerated, while
  retaining the necessary detail elsewhere.
  \sys then parallelizes bottleneck simulators by decomposing them
  into multiple parallel but synchronized processes.
  Next, \sys provides a profiler to help users understand simulation
  performance and where the bottlenecks are, so users can adjust the
  configuration.
  Finally \sys provides abstractions to make it easy for users to
  build complex large-scale simulations.
  Our evaluation demonstrates \sys in multiple large-scale case
  studies.
\end{abstract}
 \maketitle

\section{Introduction}
Research on large-scale network and distributed systems often faces the challenge of complete evaluation in a physical testbed.
Most researchers and even many practitioners do not have access to
testbeds that are large enough and/or provide the necessary flexibility,
control, and hardware.
For example, a new data center congestion control algorithm might
require specific configuration parameters at each network switch in
the data center, or a distributed system accelerated with in-network
processing requires new programmable switches deployed at specific
points in the network.

In these cases we typically rely on a patchwork evaluation that
combines end-to-end measurements in a small physical testbed with
protocol-level simulations for evaluating at scale.
However, this methodology compromises accuracy of end-to-end system
behaviors at scale.
The physical testbed is by necessity too small and protocol-level
simulations do not model many system components, from NIC behavior, to
host-interconnects, memory hierarchy, and the whole OS and
application-level software stack.

We argue that \emph{large-scale end-to-end simulation could bridge this
gap}.
In this paper, we take a top-down approach, starting from
small-scale end-to-end simulations, we tackle the practical and
fundamental challenges in scaling up.

\lijl{Can improve the transition to SimBricks.}

We use existing work on modular end-to-end simulation as a starting
point.
Modular end-to-end simulations in SimBricks~\cite{li:simbricks}
combine and connect different best-of-breed simulators for different
system components, and through modularity it can flexibly cover a broad
range of use-cases.
SimBricks scales up by running separate components as parallel
processes communicating and synchronizing through efficient
shared-memory message passing, and scales out with proxy components
that forward messages between simulator instances across hosts.
With this combination, SimBricks has been demonstrated to scale to
a simulated network of 1000 single-core hosts and NICs running
Memcached on Linux.
While technically feasible, this simulation of 10\,s of application
workload required 6--20\,h of simulation time, depending on
configuration, on 26 machines with 96 vCPUs, for \$600-\$2000 today on
ec2.

\lijl{Should we describe the challenges of scaling e2e simulation? Then we can introduce our main techniques. Otherwise, reviews might feel this is ``straightforward''}

With \sys, we enable large-scale simulations with more reasonable
cost-benefit ratios through a combination of four techniques.
\lijl{Might want to define what cost-benefit ratio mean. It's also a good opportunity to define our goals, and why those are desirable.}
First, we leverage modularity to implement \emph{mixed fidelity
simulations} where some system component instances are simulated in
less accurate simulators to drastically reduce CPU resources needed,
while key instances remain in accurate simulators.
Next, we design generic building blocks for reducing simulation time
by \emph{parallelizing bottleneck simulators by decomposing} them into
multiple parallel, connected, and synchronized processes.
We then introduce \emph{lightweight synchronization and communication
profiling} to inform the user about bottlenecks and resource
efficiency across component simulator instances.
Finally, we provide a \emph{configuration and orchestration framework}
for end-to-end simulations that simplifies specifying and running
simulations by separating the configuration of the simulated system
from concrete simulator instantiation choices.

In our evaluation we demonstrate that \sys enables evaluation of large
scale systems in networks of up to 1200 hosts, while running complete
OS and application stacks for key nodes.
\sys simulations enable full end-to-end application evaluation in
large networks.
By combining mixed fidelity simulation with parallelization through
decomposition, \sys can simulate 20 seconds in less than 4 hours while
running on a single machine.
The modular approach and flexible configuration and orchestration
framework makes \sys suitable for a broad range of evaluation
use-cases.

We plan to release \sys open source after publication.
This work does not raise any ethical issues.

\section{Background and Motivation}%
\label{sec:bg}

\subsection{Requirements}
We argue that practical simulations for evaluating large scale
systems need to satisfy the following requirements:
\begin{itemize}
  \item \textbf{End-to-End:} Obtain full end-to-end measurements
    with all relevant hardware (switches, topology, NICs,
    server-internals, etc.) and software (application, OS, library)
    components.
  \item \textbf{Scalable:} Perform measurements in systems of
    realistic scale, at least 1000s of hosts.
  \item \textbf{Efficient:} Run simulation within feasible resource
    limits, in particular processor cycles.
  \item \textbf{Fast:} Keep simulation times manageable.
  \item \textbf{Flexible:} Support simulating a broad range of
    different system configurations.
  \item \textbf{Easy to Use:} Make it easy for users to configure and
    run simulations.
\end{itemize}

\subsection{Existing Simulators Fall Short}
In this section we overview most dominant simulation tools used in network reasearch and explain why these are not enough to conduct large scale end-to-end system simulation as presented in \autoref{tab:sim_compare}. 
\textbf{Discrete Event Simulator: } DES models network behavior by sequentially processing discrete events based on their timestamp. For example a packet sent by an application or transmited to a channel, offering detailed packet-level traces and the status of the network at the certain timestamp. Established simulators like ns-3~\cite{software:ns3} and OMNet++~\cite{software:omnetinet} have garnered extensive features over decades, leading many researchers to rely on DES for evaluating proposed systems\hj{add ref}. However, due to their sequential processing nature, these simulators encounter performance bottlenecks. It often takes hours or days to simulate mere seconds. Especially when the size of the targe system increases or the number of event generated. For instance a network constitute of large number of hosts or simulating a hight bandwidth network, making simulating modern data center network with traditional DES particularly difficult.

Efforts spanning decades have aimed to parallelize DES for simulation acceleration~\cite{10.1145/84537.84545}. ns-3~\cite{software:ns3} supports distributing the network across multiple processes or physical hosts using MPI~\cite{spec:mpi} for component synchronization and communication, enhancing scalability. Nevertheless, the overhead associated with global synchronization and message passing limits potential acceleration. In our experiments, partitioning an ns-3 simulation into 16 processes on 16 processors yielded only a 3.8x acceleration. Clean-slate simulators~\cite{10.1145/166962.166979,gao2023dons,sanchez2013zsim} are designed specifically to improve the parallelism. DONS~\cite{gao2023dons} adopts a redesigned, data-oriented paradigm, optimizing cache utilization and significantly boosting simulation speed. However, they lack the extensive feature set developed over years in the field and can only simulate specific components rather than full end-to-end systems. \sys provides essential components for parallelizing existing simulators without requiring intrusive implementation changes, and a common interface to integrate all the component in the end-to-end system.

\textbf{Modular Simulator: } Simulators such as
SimBricsk~\cite{li:simbricks}, dist-gem5~\cite{mohammad:distgem5},
pd-gem5~\cite{alian:pd-gem5}, and SST~\cite{rodrigues:sst} allows users to combine multiple components such as host, hardware device, and network to construct an end-to-end simulation, and run in parallel. The main issue is the simulation speed limited by the bottleneck components. The imbalanced simulation load in each component leads to poor parallel execution. \sys presents a framework that automatically profiles each component load and indicates the balanced decomposition.

\textbf{AI Powered Estimator: } Recent efforts have delved into emulating networks using deep neural networks to estimate user-relevant metrics such as delay and packet loss. MimicNet~\cite{10.1145/3452296.3472926}, for instance, learns performance metrics at a cluster granularity level and generates estimated packet traces. DeepQueueNet~\cite{yang2022deepqueuenet} refines this approach by learning device-level performance metrics, thereby enhancing packet visibility within the cluster. These estimations are derived through inferencing input data, which lends itself well to massive parallelization, enabling rapid results for large-scale networks. However, the deep neural net's behavior is not easily interpretable, and each to model different network configurations, model has to be reconstructed, incurs significant computing and engineering effort.

\begin{table}[t]%
    \renewcommand{\arraystretch}{1.3}
    \fontsize{7pt}{10pt}\selectfont
    \begin{tabularx}{0.95\columnwidth}{ >{\centering\arraybackslash}p{1.5cm} | >{\centering\arraybackslash}X  >{\centering\arraybackslash}X  >{\centering\arraybackslash}X  >{\centering\arraybackslash}X }
    \multicolumn{1}{l|}{} & End-to-End & Scalability & Fidelity & \begin{tabular}[c]{@{}c@{}}Engineering\\ Effort\end{tabular} \\  \specialrule{1.3pt}{0pt}{1pt}
    AI Powered            &  \ding{56} &   \ding{52} & \ding{56}& High                                                              \\ 
    Original DES          &  \ding{56} &  \ding{56}  & \ding{52}& Low                                                             \\ 
    Paralle DES           &  \ding{56} &  \ding{52}  & \ding{52}& Low                                                             \\ 
    Modular Simulator    &  \ding{52}  &  \ding{56}  & \ding{52}& Low                                                             \\ 
    \sys              & \ding{52}  &  \ding{52}  & \ding{52}& Low                                                            
    \end{tabularx}
    \caption{Overview of newtwork simulators and the characteristic}
\label{tab:sim_compare}
\end{table}

\subsection{Technical Challenges}
Large-scale simulations supporting end-to-end evaluation face multiple
compounding challenges.

\paragraph{High resource needs for detailed simulators.}
In general, more detailed simulators are slower and less resource
efficient compared to less-detailed simulators.
To obtain meaningful end-to-end measurements, we generally
functionally and timing accurate simulators for all component types in the
system, be it processor and memory subsystem, hardware devices such as
NICs, and the actual network topology.
As a result, end-to-end simulations of large-scale systems are
prohibitively expensive.

\paragraph{Simulations bottlenecked by slowest component.}
Modular simulation comprising multiple synchronized components can,
by construction, only proceed as fast as the slowest component in the
system.
Slow bottleneck simulators cause two separate problems.
First, overall simulation times will be long for simulations that
include even just a single slow simulator component.
Second, as typical end-to-end simulation will naturally contain
component simulators that simulate with different speeds, this results
in faster components wasting a lot of processor cycles waiting for
slower simulators.

\paragraph{Hard to understand simulation performance.}
To make matters worse, finding bottlenecks in simulations comprising
tens to thousands of communicating and synchronized components is a
challenge.
Most efficient simulation synchronization mechanisms rely on polling
shared memory state for efficiency.
Thus all components will commonly show 100\% CPU utilization, and a
regular profiler will indicate lots of time spend in the functions
that poll for messages.
Based on these indicators it is hard to tell if a simulator is
bottlenecked or communicating heavily, especially when also combined
with heavy compiler optimization.
Blocking will also naturally propagate through dependent system
components.

\paragraph{Complex configuration and execution.}
Finally, configuring and running simulations for large-scale
end-to-end system is a complex task.
Many instances of different simulators for different components need
to be configured, connected, and then executed in a coordinated
manner.
The first problem is the complexity: each simulator has its own
mechanism and abstractions for configuring it, and there is a
substantial learning curve whenever a user looks to use a new
simulator.
Second, this is complicated by the fact that any non-trivial
evaluation typically will need to simulate multiple different
configurations of its system, and often needs to explore different
simulators and simulator configurations to identify suitable
configurations.
Finally, once the user has chosen a system and simulation
configuration, all components need to be connected together, started
in the correct order respecting dependencies, outputs need to be
collected, and finally all simulators need to be cleanly terminated.
Even with more than a handful of components a manual approach is
prohibitively complex and laborious.
\section{Design and Implementation}%
\label{sec:design}

\begin{figure}%
\centering%
\includegraphics[width=0.48\textwidth]{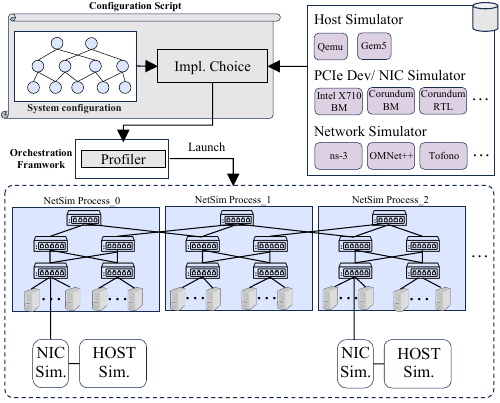}%
\caption{\sys overview}%
\label{fig:overview}%
\end{figure}

\sys combines four techniques to address these technical challenges.
\autoref{fig:overview} shows an overview.
First, \sys reduces the resources needed for large scale
simulations by enabling \emph{mixed-fidelity simulations}
(\ref{ssec:design:mixedfidelity}), where expensive detailed simulators
are replaced with faster, less resource intensive simulations in part
of the system, while keeping detailed simulation in other parts.
To increase simulation speed and avoid poorly utilized processor
cores, \sys provides generic building blocks for \emph{parallelizing
bottleneck component simulators by decomposing them into parallel
processes} (\ref{ssec:design:decomp}).
\sys helps users identify bottleneck component simulators and largely
idle component simulators with a \emph{cross-simulator synchronization
and communication profiler} (\ref{ssec:design:profile}).
Finally, \sys aims to streamline configuring and running a broad range
of different system and simulation configurations, with
\emph{programming abstractions for configuration and communication}.
(\ref{ssec:design:config}).

\subsection{Mixed-Fidelity Simulations}
\label{ssec:design:mixedfidelity}

To reduce the computational resources necessary for large-scale
end-to-end simulations, we propose mixed-fidelity simulation.
The idea is basically to retain a subset of full detailed end-to-end
components for part of the system, while using less resource-intensive
simulations for less critical areas of the system.

\begin{figure}%
\centering%
\begin{subfigure}[b]{0.22\textwidth}%
\includegraphics[width=\textwidth]{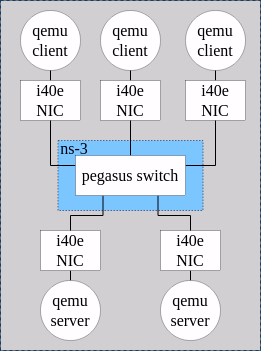}%
\caption{End-to-End}%
\label{fig:mix-fid-e2e}%
\end{subfigure} \hspace{3mm}
\begin{subfigure}[b]{0.22\textwidth}%
\includegraphics[width=\textwidth]{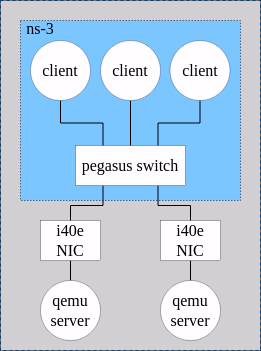}%
\caption{Mixed Fidelity}%
\label{fig:mixed-fid-hybrid}%
\end{subfigure}%
\caption{Changing an end-to-end simulation into a mixed-fidelity
  simulation by simulating clients at the protocol-level in ns-3
  instead using individual host and NIC simulator instances.}%
\label{fig:mix-fid-example}%
\end{figure}

\paragraph{Reducing Simulation Detail in non-Critical Components.}
The underlying insight is that typically full detail is not required
in every component of the system.
A common example is running a system as part of a larger network to
evaluate the effect of other background traffic, congestion, etc. in
the network; here protocol-level simulation of hosts generating this
background traffic is completely sufficient.
However, where detailed simulation is required and where detail can be
sacrificed depends on the system and evaluation goal.
When evaluating peak system throughput for a client-server system,
modelling internal client detail is not essential --- as long as
client requests arrive at the required rate and with the correct
protocol or format, the server behavior will be the same.
When evaluating end-to-end request latency, on the other hand, client
internal behavior is likely to significantly affect measured latency,
thus here at least the clients that measure the latency need to be
simulated in full detail.
For all three of these examples, instead of simulating all hosts
end-to-end with detailed architectural simulators, such as qemu or
gem5, we can instead simulate a specific subset of them at the
protocol level, e.g. in ns-3 or OMNet++
(\autoref{fig:mix-fid-example}).
A similar approach applies to other system components, e.g. instead of
running expensive RTL-level simulations for all NICs or Switches for
projects that propose new hardware design, a judicious combination of
RTL simulations with faster and more efficient lightweight simulation
models, drastically reduces cycles needed.

\paragraph{Enabling Mixed-Fidelity End-to-End Simulations.}
At a technical level, \sys enables mixed fidelity simulations through
modular composition, inherited from SimBricks.
Components simulators are connected through fixed message passing
interfaces, primarily Ethernet packets and PCI, and individual
simulators are thus decoupled from how these messages are generated.

\paragraph{New Challenges.}
However, while mixed fidelity simulations can drastically reduce
computational cost for large-scale simulations, configuring and
running mixed fidelity simulations gives rise to or exacerbates the
other three challenges.
First, these simulations often result in heavy bottlenecks for
simulation speed, thereby also introducing significant imbalance
leading other simulators to waste cycles waiting and leaving cores
idle --- with most SimBricks simulations we have run, the end-host
simulators (qemu or gem5) are the slowest component by a significant
margin, however once we move a few hundred or thousand hosts into the
ns-3 network, ns-3 slows down the whole simulation by 3--5$\times$.
In the following two subsections we discuss how \sys enables users to
locate (\ref{ssec:design:profile}) and mitigate
(\ref{ssec:design:decomp}) such bottlenecks.
Finally, configuring mixed-fidelity simulations and exploring
different levels of detail in different system components, is
particularly complicated and laborious for users.
In addition to building host disk images with applications and
configurations, and setting up commands for each host to run etc., a
mixed fidelity simulation now also requires configuring additional
simulators, e.g. ns-3, to also implement similar functionality through
their abstractions.
\sys simplifies this, in part, through the configuration and
orchestration framework (\ref{ssec:design:config}).

\subsection{Parallelizing Through Decomposition}
\label{ssec:design:decomp}

In general, parallelizing simulators is a challenging problem with
different approaches for different types of simulators.
These are well-studied but at least the few major relevant simulators
for end-to-end simulations are either sequential (gem5) or scale
poorly (ns-3, OMNeT++)~\cite{zhang:mimicnet}.
Moreover, the existing parallelization approaches often require
intrusive changes to simulators.
In \sys we instead propose simpler easy-to-integrate building blocks
to parallelize system simulators with modular architectures, such as
ns-3, OMNeT++, and gem5.

The key idea in the \sys parallelization approach is to decompose
these simulators at component boundaries into multiple separate
processes.

We then leverage the well-defined module interfaces for connecting and
synchronizing the parallel processes with \sys adapters that translate
these interface into messages on SimBricks channels, the same channels
also used to interconnect other \sys simulator components.
Using the same mechanisms enables re-use and provides \sys with
visibility into the simulation structure for effective orchestration
and also enables use the \sys profiler for these newly parallel
components.

\subsubsection{Building Blocks}

\paragraph{Base adapter.}
We build on SimBricks adapters in gem5, ns-3, and OMNeT, that are all
implemented simply within the device abstractions of each simulator,
and implement synchronization through the channel, as well as
communication.
Based on these, we define an abstract \sys base adapter for each
simulator, that implements initialization and synchronization, but is
not specific to a particular SimBricks channel type.
This base adapter can then be used to implement multiple specific protocol
adapters without needing to re-implement the common functionality.
This includes adapters for the existing SimBricks protocols, but also
makes it easy to implement adapters for internally connecting and
synchronizing pieces of a simulator.

\paragraph{Trunk adapter}
Many non-trivial partitions will require multiple connections between
some pairs of processes.
In principle here multiple instances of the \sys adapter can be used
and this will just work.
However, this will unnecessarily incur the synchronization overhead
once for each adapter.
To address this, \sys introduces trunk channels, that multiplex
messages for multiple upper layer channels over one synchronized
SimBricks channel.
The implementation tags messages going across with the sub-channel
identifier, for demultiplexing at the receiver.

\subsubsection{Examples}
\begin{figure}%
\centering%
\begin{subfigure}[b]{0.22\textwidth}%
\includegraphics[width=\textwidth]{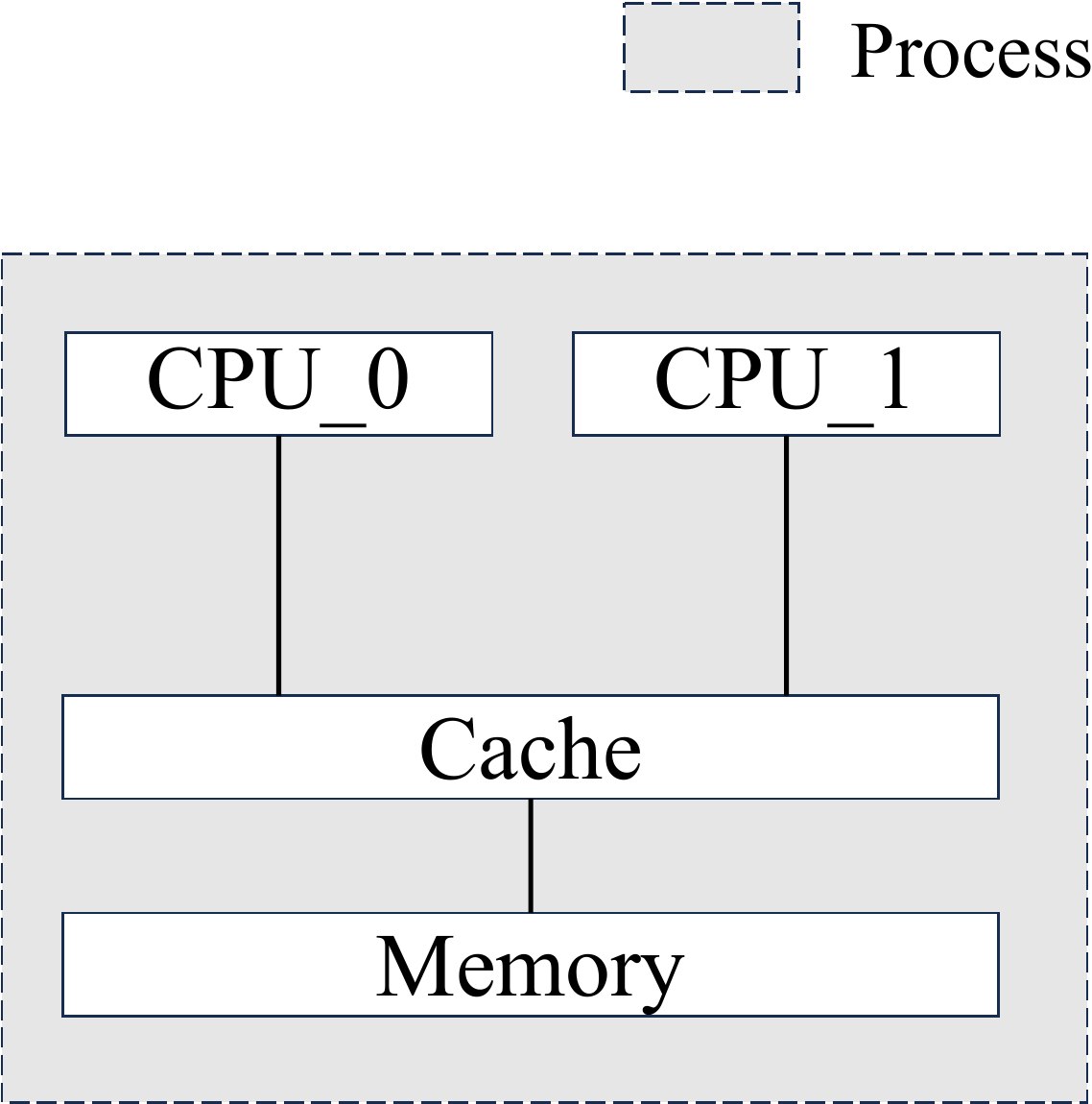}%
\caption{gem5 single-process}%
\label{fig:gem5-single-core}%
\end{subfigure} \hspace{3mm}
\begin{subfigure}[b]{0.22\textwidth}%
\includegraphics[width=\textwidth]{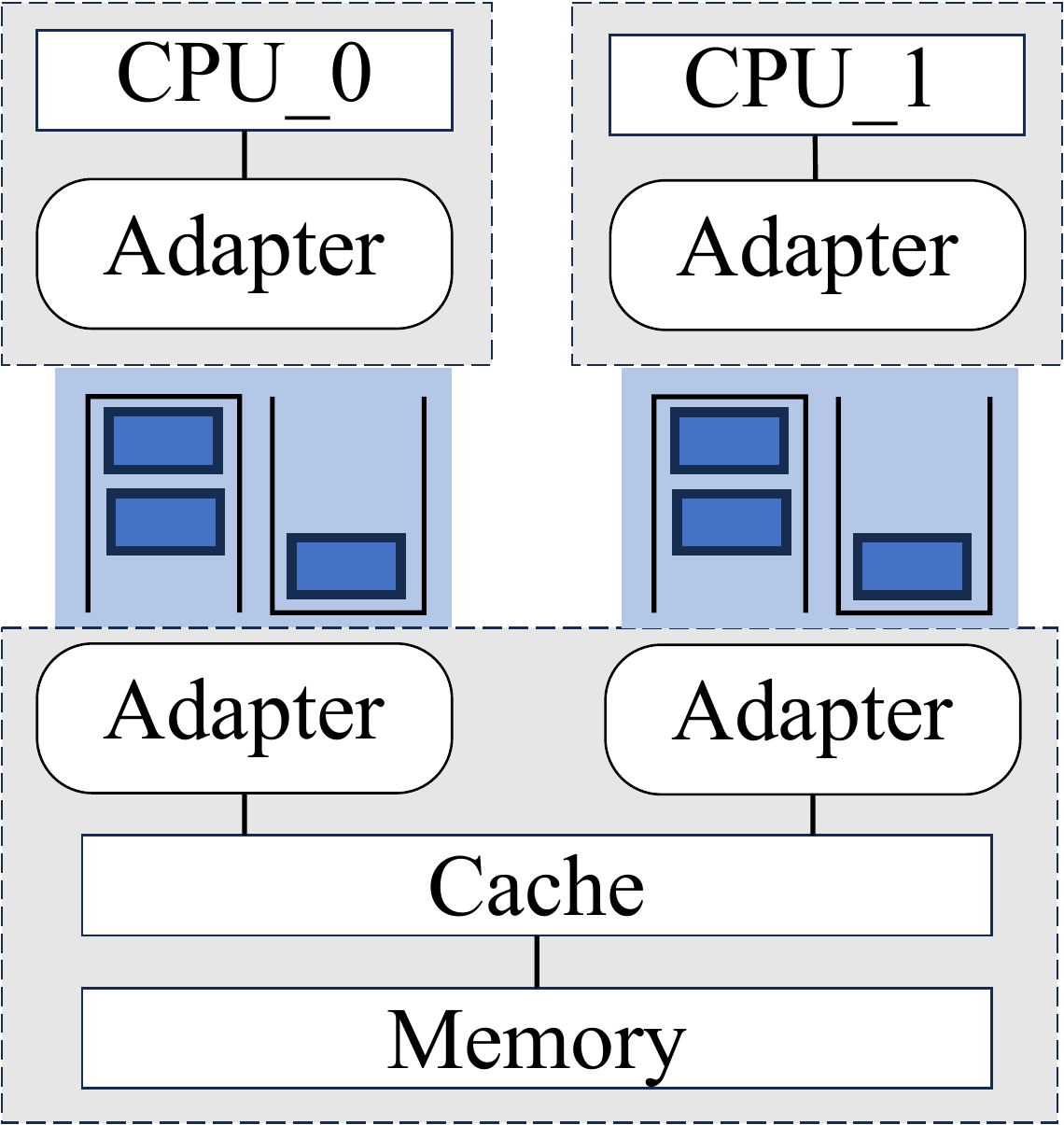}%
\caption{gem5 multi-process}%
\label{fig:gem5-multi-core}%
\end{subfigure}%
\caption{Parallelizing a sequential multicore architecture simulation
  by splitting it into  parallel processes interconnected with \sys
  adapters.}%
\label{fig:gem5-multicore-figure}%
\end{figure}

\paragraph{Multi-Core gem5.}
\autoref{fig:gem5-multicore-figure} demonstrates how we use \sys
adapters to parallelize multi-core simulations in gem5.
Changes to gem5 are limited to 1) implementing the adapters as
simulation object in gem5, and serializing the already message-based
memory packet interface to messages, and 2) changing the gem5 python
configuration script to only instantiate the relevant components for
each process.

\paragraph{Parallel ns-3 and OMNeT++.}
We also implemented \sys parallelization for the ns-3 and OMNet++
network simulators.
\todo{figure; should match figure for ac profiler example in eval}
Here we also instantiate different parts of the overall network
topology in separate processes, and replace links going across
components with \sys trunk link adapters.
For network simulators we rely on the user to configure the
partitioning and create the adapters, either manually or through our
configuration framework (\ref{ssec:design:config}).

\subsection{Lightweight Profiling for Synchronization and Communication}
\label{ssec:design:profile}

To address the challenges in understanding simulation performance,
deciding what to parallelize, consolidate, and generally find
bottlenecks, \sys includes profiling infrastructure.
The \sys profiler measures metrics related to \sys cross-simulator
synchronization and communication in each component simulator.
The profiler comprises two components: instrumentation in each
simulator, and post-processing to aggregate the collected metrics and
present them to the user.

\subsubsection{Lightweight Instrumentation}
\sys instruments each adapter, both for communication across simulator
components and within the processes of a particular component, with
lightweight metric collection and logging.
First, each adapter continously counts the number of 1) \emph{CPU
cycles blocked waiting for a synchronization message} from the peer to
allow the simulation to proceed, 2) \emph{sending data messages} to
peer simulators, and 3) \emph{processing incoming data messages}.
Second, each simulator can be set to periodically, e.g. every 10\,s,
log the values of these counters for each adapter and the current time
stamp counter as well as that simulator's current simulation time.

\subsubsection{Profiler Post-Processing}
After the simulation terminates, either because it completes or
because the user stops it, the profiler post processor ingests and
parses these logs.
As each simulator logs absolute totals for each value, we calculate
the difference between a late entry towards the end and an early entry
towards the beginning, dropping a configurable number of warm-up and
cool-down lines.
Each log entry contains both the simulation time and processor time
stamp counter, thereby providing a reference for simulation time and
physical system time.

\paragraph{Metrics calculated}
The post processor first calculates a global metric, \emph{simulation
speed}, by dividing the difference in simulation time by the
difference in time stamp counter cycles (as all simulators are
synchronized, this value is the same for each simulator).
For each simulator we also calculate their \emph{efficiency} as the
fraction of cycles not spent on receive, transmit, or synchronization
in the \sys adapter.
This metric is useful to determine when diminishing returns for
parallelizing \sys simulations set in.

\begin{figure}%
\centering%
\includegraphics[scale=0.5]{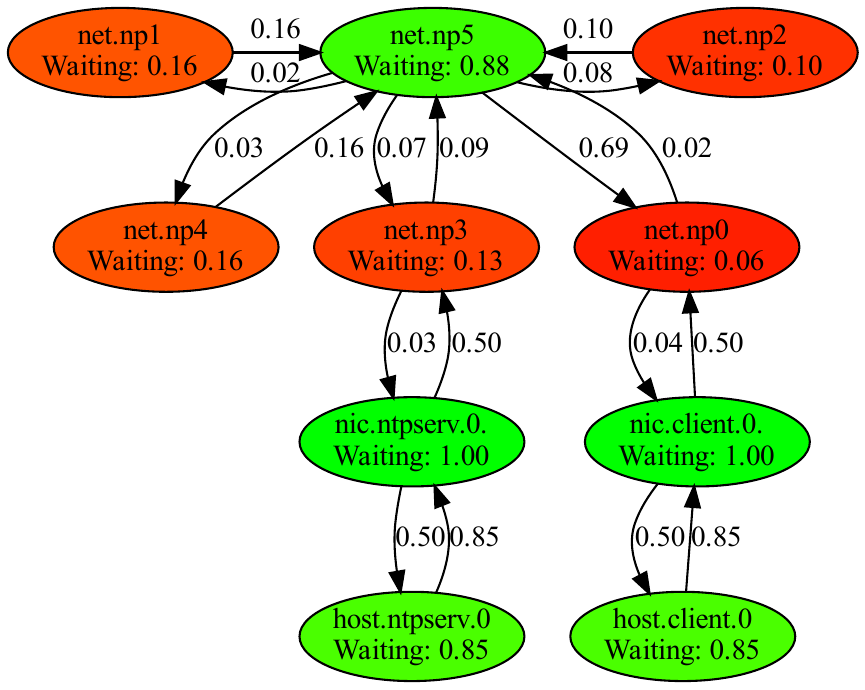}%
\caption{Example of a generated wait-time-profile graph. Here the
  \texttt{net.np0} process is the immediate bottleneck, but
  \texttt{np1-3} are close behind, judging from their waiting
  numbers.}%
\label{fig:profile-example}%
\end{figure}

\paragraph{Wait-Time Profile Graph}
The main output for understanding \sys simulation performance and for
localizing bottlenecks, is the wait-time profile graph (WTPG).
The WTPG contains a node for each simulator instance, and a pair of
opposite directed edges for each \sys channel connecting two
simulators.
The profiler annotates each edge with the fraction of cycles that the
simulator at the source of the edge has spent waiting for
synchronization messages from the destination simulator of the edge.
As such, the graph visualizes "who waits for who".
Additionally, the profiler annotates each node with the total number
of cycles that node spends waiting across simulators.
Based on this value, we also color nodes on a spectrum from green to
red, with red for nodes that spend few cycles waiting for other
nodes, and green for nodes that spend many cycles waiting for other
nodes.
Typically, nodes that spend little time waiting, are the bottleneck
simulators and will stand out in red.
If in doubt, the edge labels allow users to confirm that their
neighbors spend significant cycles waiting on them.
\autoref{fig:profile-example} shows an example of a WTPG for a \sys
simulation.

\subsection{Configuration and Orchestration}
\label{ssec:design:config}
Finally, we are left with addressing the complexity of configuring and
running a broad range of large-scale simulations.
\sys addresses this with an orchestration framework.
The orchestration framework aims to reduce the user configuration
complexity by providing natural abstractions for separately specifying
the configuration of the simulated system from the implementation
choices for how to simulate the system.
Finally \sys will apply the specified implementation choices and
coordinate the execution of the simulation, including starting up each
component simulator, connecting them up, collecting outputs, and,
eventually, clean up after termination.

Crucially, with the \sys configuration abstractions, users can solve
many simulation configuration tasks fully within \sys, without
resorting to manually configuring specific simulators through their
specific configuration mechanism.
For such tasks, \sys abstractions also provide a level of portability,
in fully separating system configurations from concrete simulator
choices.
At the same time, the \sys orchestration framework can easily be
bypassed where necessary and users can resort to manually configuring
specific simulators.
The \sys orchestration aims to make easy tasks easy, and complex tasks
possible.

\subsubsection{System Configuration Abstraction}
The goal for the \sys system configuration abstraction is to specify the
\emph{configuration of the simulated system} separate from concrete
choices of how to simulate it.
We represent the system configuration as a hierarchy of Python
objects.
At the root we have the \texttt{SystemConfiguration} object, that
contains a list of all system components.
A system component can be a host, a NIC, a switch, or a PCI or
Ethernet link.
Each component object carries the expected attributes.
For example, a host object may specify the number of cores, memory,
disk image, applications to run, IP address, etc.
A link object specifies latency and bandwidth, along with its two
endpoints.
The key design consideration for these abstraction is to specify
system characteristics while abstracting simulation details.

Since we use Python for defining \sys system configurations, users can
use all python language features for assembling this configuration, in
particular relying on loops for instantiating repeated patterns or
using functions and modules for factoring out re-usable configuration
parts.
For example, for the experiments in this paper we use the same
parametrizable large-scale network topology across multiple of our
experiments and have defined this in a common function across
experiments.

\subsubsection{Implementation Choices}
After a user has assembled a system configuration in the simulation
Python script, the second step is to generate one or more different
concrete simulation instantiations.
Users instantiate a \sys simulation by choosing specific simulators
and translating the system configuration for the corresponding system
components into configurations for these concrete simulators.
We specify the resulting instantiated simulation using the existing
SimBricks abstractions for describing interconnected instances of
component simulators.

In general, there are many different instantiation strategies.
Instead of trying to automate this inherently complex step with a
one-size-fits-all approach, \sys instead opts for an extensible and
flexible approach by merely providing library routines for common
instantiation strategies.
For example, one strategy we commonly use is to instantiate all hosts
as separate processes for a specfied host simulator, qemu or gem5, all
NICs of a particular type, and simulate the whole network topology in
one ns-3 process.
A generalized version of this strategy instead first applies a
partition function provided as a parameter to divvy up the network
topology components into different partitions to run in separate ns-3
processes.
For the network topology above, we have implemented a couple of
different partition strategy functions that we use across the
experiments in this paper.

As instantiated simulation configurations is just a regular SimBricks
configuration, comprising SimBricks orchestration python objects, \sys
users can manually modify this configuration afterwards when the
need arises.

\subsubsection{Running Simulations}
Finally, to actually run \sys simulations, we leverage the existing
SimBricks orchestration framework runtime.
Since the instantiation above produces a SimBricks configuration as
its output, this can directly be passed through for execution.
\section{Evaluation}%
\label{sec:eval}

In our evaluation we aim to answer the following questions:

\begin{itemize}
  \item Does \sys enable \emph{end-to-end} evaluation of networked
    systems that are hard to evaluate in physical testbeds?
  \item What are the resource savings from mixed-fidelity simulations?
  \item How much accuracy do mixed-fidelity simulations sacrifice
    compared to full end-to-end simulations?
  \item How effective is \sys parallelization at removing simulation
    bottlenecks? Is it competitive with native parallelization
    approaches in specific simulators?
  \item Is simulation speed for different parallelization strategies
    hard to predict and understand? Does the \sys profiler help?
  \item What user-effort is required to configure and run \sys
    simulations?
\end{itemize}

\subsection{Methodology}
Our measurements are performed on machines with double Intel Xeon Gold
6336Y CPUs for a total of 48 physical cores and 256GB RAM.
For resource efficiency, we opt for the smallest simulations that
substantiate each evaluation point.
While \sys supports SimBricks proxies for distributed simulations and
inherits their demonstrated scalability, by relying on mixed fidelity
simulations for our evaluation we have not found the need to scale out
to multiple machines.

Unless otherwise mentioned, we use the following simulation
parameters.
We configure up single-core hosts and qemu with instruction counting
for time synchronization.
Simulated hosts are configured with 4\,GHz clock frequency and 1\,GB
of memory.
For NICs we use the SimBricks \texttt{i40e\_bm} simulator for the
Intel X710 NIC.

\subsection{Case-Study: In-Network Processing}
We start off with a case study on evaluating
NetCache~\cite{jin:netcache} and Pegasus~\cite{li:pegasus}, two
distributed storage systems with in-network support.
NetCache caches key-value items in programmable switches but directs
writes to a single responsible replica.
Pegasus instead load-balances all requests to servers but uses the
switches as an in-network coherence directory to enable load-balancing
writes to multiple hosts.
For this, we implement the switch functionality for both systems in
ns-3.
In all configurations we simulate two servers and three clients
connected to a single switch, and configure the client with a skewed
zipf 1.8 key distribution and 70\% write workload.
For the protocol-level simulation, we implement both the client and
server as ns-3 applications, while the mixed-fidelity simulation only
uses the former.
The end-to-end and mixed-fidelity simulations use the unmodified
client and server Linux applications from the project repositories.

\begin{figure}%
\centering%
\includegraphics[width=0.4\textwidth]{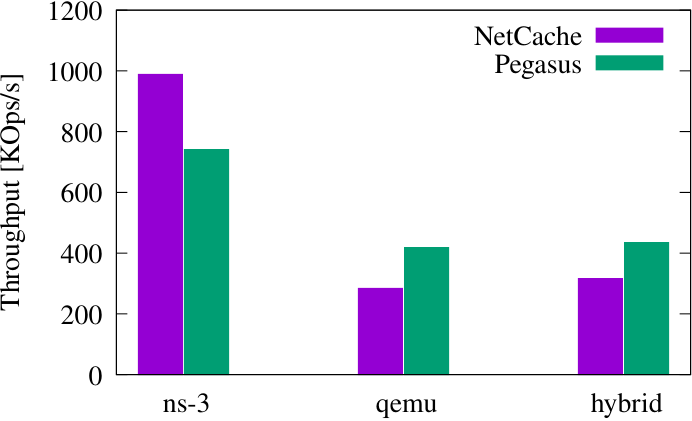}%
\caption{Comparing NetCache and Pegasus throughput, with different
  simulation configurations.}%
\label{fig:nc_pegasus_cmp}%
\end{figure}

\paragraph{Need for End-to-End Simulation.}
First, we compare results from protocol-level ns-3 simulation
to a full end-to-end simulation and a mixed-fidelity simulation.
We configure all clients with the same offered load.
We start by measuring system throughput, and report results in the
first two groups of bars in \autoref{fig:nc_pegasus_cmp}.
This comparison shows opposite trends, the protocol-level ns-3
simulation shows NetCache outperforming Pegasus by 33\%, while the
end-to-end simulation shows 47\% higher throughput for Pegasus.
Inspection of simulation logs shows that this is because the
end-to-end system is bottlenecked by server software process, which
ns-3 does not model.
This also shows up in latency measurements for this workload, even
under saturation the request latency in ns-3 comes out to 7--8$\mu$s,
while the end-to-end simulation measures 590--704$\mu$s.
This demonstrates the need for end-to-end measurements.

\paragraph{Benefit of Mixed-Fidelity Simulation.}
However, the end-to-end simulation uses 11 cores (1 host simulator and
1 NIC simulator per server, plus ns-3 instance for network), while the
protocol-level simulation only needs one core.
On top of this, the end-to-end simulation also needs these cores for a
complete simulation time of 1160\,s, while the protocol-level simulation
only needs the core for 170\,s.

Given that this setup saturates the servers and measuring system
throughput, we primarily need detailed simulation behavior for the
servers, while client internals do not affect end-to-end performance
significantly.
Thus, we configure a mixed-fidelity simulation that only simulates
servers with separate host and NIC simulators, while simulating
clients in ns-3.
As a result, this configuration needs 54\% fewer cores, with 5 total,
2 host simulators and NICs for the servers plus one core for ns-3, and
has a 17\% lower simulation time.
The rightmost group of bars in \autoref{fig:nc_pegasus_cmp} shows
similar throughput for the mixed-fidelity simulation.

\begin{figure}%
\centering%
\begin{subfigure}{.24\textwidth}%
\centering%
\includegraphics[scale=.75]{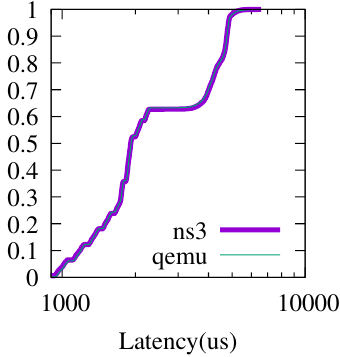}%
\caption{Saturated Servers}%
\label{fig:pegasus_lat_sat}%
\end{subfigure}%
\begin{subfigure}{.24\textwidth}%
\centering%
\includegraphics[scale=.75]{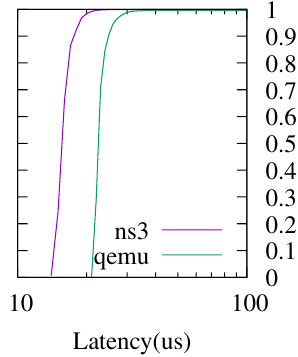}%
\caption{Un-saturated Servers}%
\label{fig:pegasus_lat_unsat}%
\end{subfigure}%
\caption{Pegasus latency CDFs for ns3 client and qemu client in two
  different mixed-fidelity simulations, one saturating the servers,
  and one with low request load.}
\label{fig:pagasus_latency}%
\end{figure}

\paragraph{Latency in Mixed-Fidelity Simulation.}
When it comes to measuring latency, the situation is less clear cut,
here client behavior may or may not be a significant factor.
We demonstrate this in a mixed fidelity simulation that replaces one
ns-3 clients above with a QEMU host and NIC simulator, and measure
latencies from the ns-3 clients as well as the qemu client.
We compare two different client workloads, first the one above that
saturates the servers, and then a low-throughput workload that does
not saturate the servers.

\autoref{fig:pagasus_latency} shows latency CDFs for both workloads.
In \autoref{fig:pegasus_lat_sat} we see that under saturation, when
latencies are in the milliseconds, both ns-3 and qemu clients measure
the same latency distribution, as the client contribution here is
negligible.
For the low-throughput workload in \autoref{fig:pegasus_lat_unsat}
latencies are significantly lower, leading ns-3 and qemu clients to
measure significantly different latency distributions.

\subsection{Case-Study: Clock Synchronization}
\label{ssec:eval:clocksync}
In our second case study, we aim to compare NTP vs PTP host clock
synchronization accuracy and its effect on application performance for
distributed systems that rely on clock-bounds for
consistency~\cite{meta-blog,corbett:spanner} in large-scale networks.
We use a modified version of CockroachDB~\cite{lindfors:cockroach}
that uses the dynamic clock bound from the Chrony NTP
server~\cite{chrony} for it's commit-wait period, used in prior
work~\cite{gunnarsson:clockdistdb}.
We configure the following end-to-end host machines: 2 CockroachDB
replica servers, 4 CockroachDB clients running the \texttt{social}
workload, and a single clock server, either NTP server or PTP grand
master.
Clients and server run Chrony, for PTP alongside \texttt{ptp4l}.
For the NTP configuration, we configure Chrony to synchronize to the
NTP server.
For the PTP configuration, we configure Chrony to use the local NIC's
PTP hardware clock (PHC) as a reference clock.

We integrate these machines into a large-scale network topology
comprising 1200 hosts total, 7 qemu hosts, and 1193 background hosts
simulated in NS3.
The background hosts are randomized pairs of hosts performing bulk
transfers.
The network topology is organized as a single core switch, connected
through 100\,Gbps links to 4 aggregation switches, that each connect
to 6 racks with a ToR and 40 machines.
We extended ns-3 with a switch that implements a PTP transparent clock
(TC).

In the simulation, we measure the clock accuracy bound that Chrony
reports on the servers.
As expected, we see that PTP, with its NIC hardware timestamping and
transparent clocks in switches, improves the clock bound from
11\,$\mu$s
with NTP, to 943\,ns with PTP.
Note that this includes a full end-to-end simulation of the PTP
synchronization, with \texttt{ptp4l} running on Linux, using the NIC
hardware receive and transmit timestamping, as well as the transparent
clock switch support adding corrections for queue residence time.
As reported by prior work~\cite{meta-blog,gunnarsson:clockdistdb},
this improved clock bound improves application request throughput and
latency for the application relying on commit-wait to ensure
consistency.
We measure a 38\% throughput improvement for write operations, and a
15\% reduction in latency for writes.
This simulation simulates 20s in 175\,min and 227\,min for NTP and PTP
respectively.

\subsection{Case-Study: Congestion Control}
\begin{figure}%
\centering%
\includegraphics[width=0.4\textwidth]{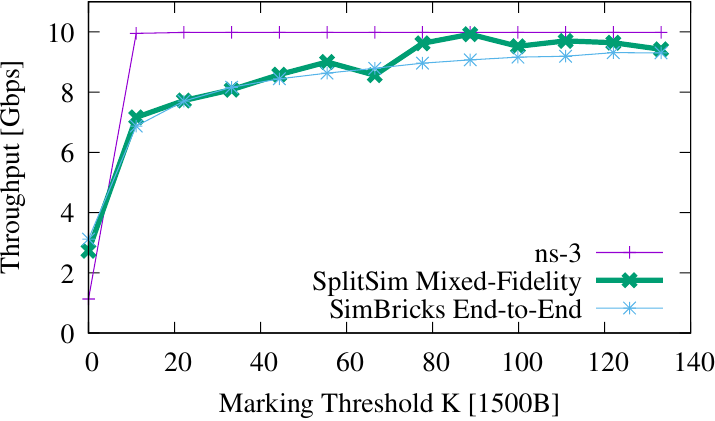}%
\caption{DCTCP congestion control behavior with different marking
  threshold, evaluated in ns-3, a mixed-fidelity, and end-to-end
  simulations.}%
\label{fig:hybrid-dctcp-dumbbell}%
\end{figure}

In a final case study, we evaluate the suitability of mixed-fidelity
simulations for evaluating congestion control implementations.
As network bandwidths have increased and latencies decreased,
host-internal behavior, such as processing time variation or other
bottlenecks, increasingly affect congestion control
behavior~\cite{mittal:timely,agarwal:hicongestion,kumar:swift}.
None of these behaviors are modelled by the common protocol-level
simulations for congestion control algorithms, reducing the ability
of these simulation results to predict real system behavior.
Prior work~\cite{li:simbricks} has demonstrated that end-to-end
simulation can improve validity of the results.
But as congestion control schemes, especially for data centers,
typically need to be evaluated at scale, the resulting simulation
resource requirements are prohibitive.

To address this, we now evaluate whether mixed-fidelity simulations
can provide sufficiently accurate results relative to full end-to-end
simulations.
To that end, we simulate a common typical dumbbell topology with a 10G
bottleneck link and two hosts on each side, performing a bulk transfer
across the bottleneck link.
We use dctcp~\cite{alizadeh:dctcp} congestion control, and evaluate
the effect of different marking threshold parameter values on
throughput.
We compare three cases: 1) protocol-level simulation in ns-3, 2)
mixed-fidelity simulation with one pair of gem5 hosts, and one pair of
ns-3 hosts, and 3) full end-to-end setup with all 4 gem5 hosts.
autoref{fig:hybrid-dctcp-dumbbell} shows that the mixed fidelity
simulation behavior closely matches the end-to-end simulation, while
protocol-level simulation is far off.

\subsection{Parallelizing Simulators with \sys}
Next, we evaluate our \sys simulator parallelization approach relying
on decomposition and SimBricks synchronization and communication.
First, for parallelizing the sequential gem5 simulator, and then we
compare \sys parallelization to native parallelization mechanisms in
ns-3 and OMNeT++.

\subsubsection{Parallelizing Sequential gem5 Multi-Core Simulations.}
The gem5~\cite{binkert:gem5} architectural simulator is sequential.
As a consequence, when simulating a multi-core machine, simulation
time increases at least linearly with the number of simulated cores.
Simulating larger multi-core hosts is prohibitively slow.
At the same time, gem5 takes a similarly modular configuration
approach based on standard component interfaces, similar to network
simulators such as ns-3 or OMNeT++.
Specifically, gem5 components, such as processor cores, caches,
memories, or devices, connect through ports that communicate through
packetized memory requests.

\begin{figure}%
\centering%
\includegraphics[width=0.4\textwidth]{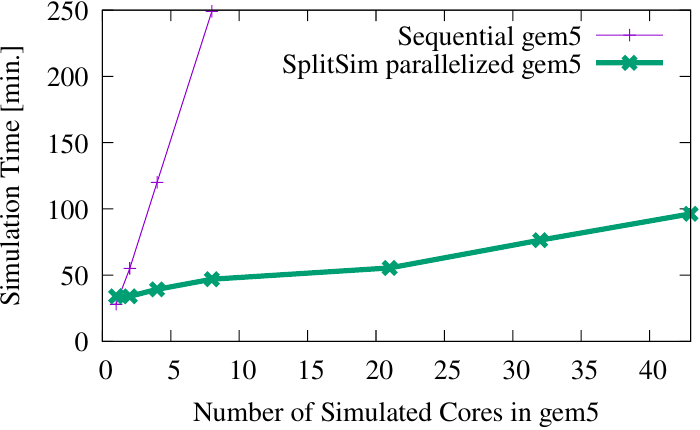}%
\caption{Simulation time for \sys-parallelized multi-core gem5
  compared to sequential gem5.}%
\label{fig:gem5-multicore}%
\end{figure}

We leverage this and implement a \sys adapter for this interface, that
forwards these messages across a SimBricks channel to a different
process.
This required roughly 1000\,LoC of code to be added to gem5, without
intrusive changes required.
We then create separate gem5 configurations to simulate each processor
core in parallel as a separate process, connected together through
\sys channels.
We validate through detailed simulator logs with timestamps that the
parallelized multi-core simulation behaves as the original sequential
simulation.
\autoref{fig:gem5-multicore} shows drastically reduced simulation time
of the parallelized simulation compared to sequential gem5.
For 8 cores, we see about a 5$\times$ speedup.
The parallelized gem5 also scales well, with simulation time only
increasing by a factor of 2 from 8 cores to simulating 44 cores.

\subsubsection{\sys vs. native parallelization in ns-3 and OMNeT++.}
\begin{figure}%
\centering%
\includegraphics[scale=.65]{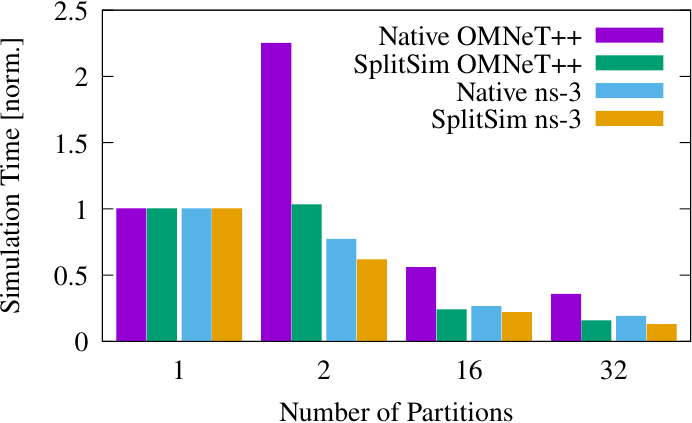}%
\caption{Comparing \sys parallelization to native parallelization in
  omnet and ns-3.}%
\label{fig:net_par}%
\end{figure}

Next, we compare \sys parallelization for the ns-3 and OMNeT++ to
their native MPI-based parallelization schemes.
For this we leverage the SimBricks network adapters already
implemented in these simulators for connecting to NIC simulators.
We use the FatTree8 network configuration from
DONS~\cite{gao2023dons}, comprising 128 simulated servers.
We then evenly partition this topology into 1, 2, 16, and 32,
components, once with the native synchronization mechanism and once
with \sys.
\autoref{fig:net_par} shows that \sys outperforms both ns-3 and
OMNeT++ synchonization, resulting in up to 57\% lower simulation
times.
Despite the lightweight integration into simulators, \sys vastly
outperforms native parallelization for both simulators.

\subsection{\sys Profiler}
\begin{figure}%
\centering%
\begin{tabular}{cp{2.5in}}%
\toprule
\textbf{Part.} & \textbf{Description} \\
\midrule
  s   & Whole network as one process. \\
  ac  & One process per aggregation block, plus one for the core
        switch. \\
  crN & Aggregate N racks into a process, plus one for the aggregation
        and core switches. \\
  rs &  One process per rack, one process each per aggregation and core
        switch.\\
\bottomrule
        \\
\end{tabular}
\includegraphics[width=0.4\textwidth]{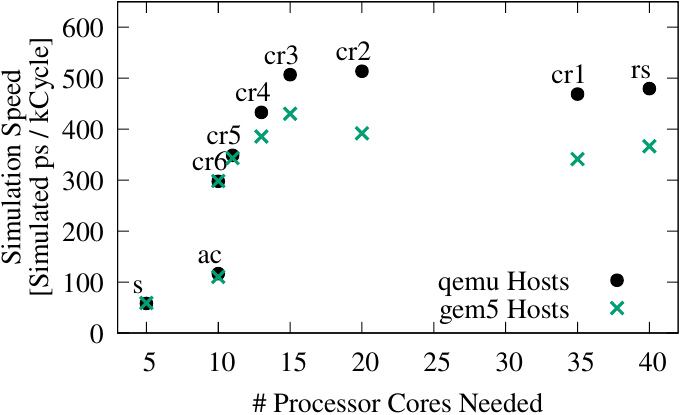}%
\caption{Simulation speeds for different network partition partition
  strategies with qemu and gem5 hosts.}%
\label{fig:ns3-part-strat-perf}%
\end{figure}

We now demonstrate the challenges in predicting simulation performance
a priori, and demonstrate demonstrate using the \sys profiler to find
simulation bottlenecks.

\paragraph{Complexity in Predicting Simulation Performance.}
We first explore different partitioning strategies for the 1\,200 node
network topology with background traffic from
\autoref{ssec:eval:clocksync}.
Here we connect a pair of qemu or gem5 hosts with two Intel x710 NICs.
\autoref{fig:ns3-part-strat-perf} shows the partition strategies along
with their achieved simulation speeds and cores used for the whole
simulation (including 4 cores for hosts and NICs).
The results show that different partitioning strategies achieve
significantly different simulation speeds across strategies but also
with qemu compared to gem5, in some cases even with identical number
of cores.
Additionally, the results show that past a point adding more cores
results in lower simulation speeds again.

\begin{figure}%
\centering%
\begin{subfigure}{0.45\textwidth}%
\centering%
\includegraphics[scale=0.32]{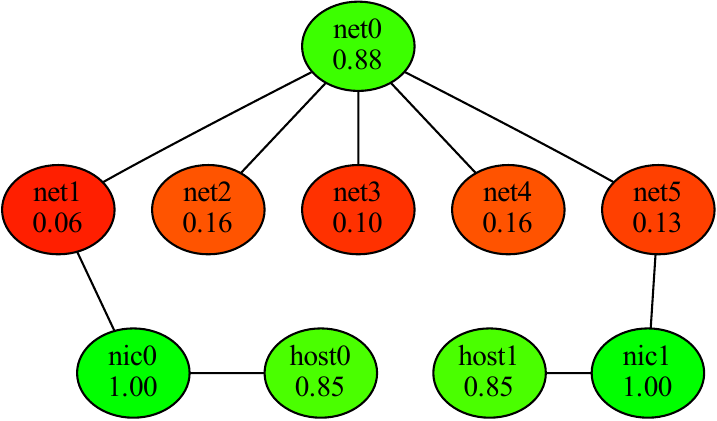}%
\caption{ac partition strategy}%
  \label{fig:part_prof_ac}%
\end{subfigure}
\begin{subfigure}{0.45\textwidth}%
\centering%
\includegraphics[scale=0.32]{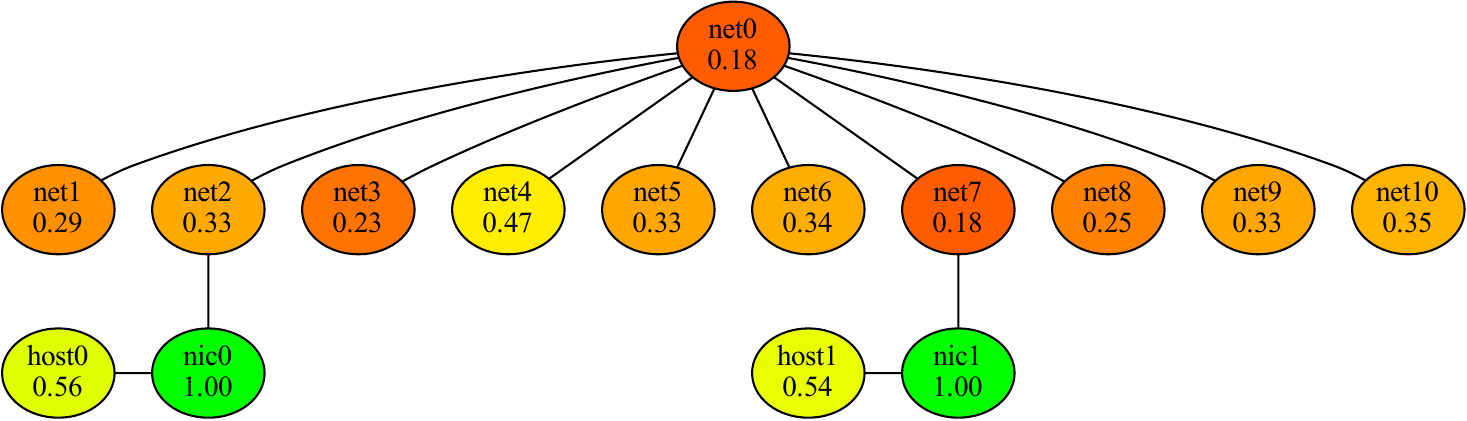}%
\caption{cr3 partition strategy}%
\label{fig:part_prof_cr3}%
\end{subfigure}%
  \caption{\sys profile graphs for two partition strategies in
  \autoref{fig:ns3-part-strat-perf} with qemu hosts.}%
\label{fig:part_prof}
\end{figure}

\paragraph{Profiling to Locate Bottlenecks.}
Next we pick the \emph{ac} and \emph{cr3} partition strategies and
examine their \sys profile graphs in \autoref{fig:part_prof}.
Here, we run the simulations for 5\,min for the most reliable results,
but we found typically even 60-90\,s is sufficient to record the
profile to inform partition decisions.
Profiling and post-processing to generate the graphs are fully
automatic, and simply require adding the flag to enable profiling when
running \sys, and then running the post-processing script.
For compactness and readability, we simplify the graph here to only
show the node color representing the fraction of cycles each simulator
spends waiting for messages in the \sys adapters.

High waiting cycles, shown in green, imply the simulator is not
computation bound and thus not the bottleneck, while few waiting
cycles, shown in red, imply a bottleneck.
\autoref{fig:part_prof_ac} shows that for the coarse-grain
\texttt{ac},
primary bottlenecks are the ns-3 instances with the 6 racks, rather
than the ns-3 instance with the core, or the qemu or NIC instances.
Thus, as expected, a much more fine-grain partition of the network
into 15 processes with \texttt{cr3}, \autoref{fig:part_prof_cr3} shows
that here the bottleneck are starting to shift towards the two qemu
instances.

\antoine{Could add another quick experiment with the profiler overhead,
just run the same again with profiling compiled out of all adapters.
we have \#defines for this}

\subsection{\sys Config and Orchestration}
Finally, we demonstrate the flexibility and ease of use of \sys, by
comparing the necessary effort for configuring and running some of the
simulations used in our evaluation.

\paragraph{\sys Simulations are Easy to Configure.}
Even complex simulation configurations are relatively easy to
configure in \sys.
For example, the configuration for running all the clock
synchronization simulations in \autoref{ssec:eval:clocksync} comprises
252 lines of Python, 195 of which are responsible for generating
configuration files and commands to run for Chrony, ptp4l, and
CockroachDB.
The other simulation configurations in this evaluation section are
more compact.
Other than the initial extension to ns-3 for PTP transparent clocks in
switches, no changes outside of the python configuration are required
for configuring these simulations.

\paragraph{Re-use of Configuration through Python.}
\sys configurations are just Python scripts; ordinary language
features such as loops, functions, and modules, can be used as
meta-programming for generating configurations.
We commonly use loops to generate different structurally similar,
configurations.
We also abstract out common building blocks into re-usable python modules.
For example, the large (parametrized) background network topology used
in multiple experiments is defined in a separate python module of
195 lines and is imported and used for multiple of our simulations.

\paragraph{Running Simulations is Fully Automatic.}
After writing a \sys simulation configuration, execution, is fully
automatic.
The \sys orchestration framework starts processes, wires up channels
between different simulator instances, collects output, and cleanly
terminates simulations.

\section{Related Work}%
\label{sec:related}
\textbf{Discrete Event Simulator: }Discrete Event Simulators (DES) stand as the cornerstone in network systems research. ~\cite{lantz:mininet,riley2010ns,varga2010overview}. While existing DES excel in capturing network behavior intricacies at the packet level, their sequential nature impedes scalability as simulation time fails to align with the target system's size. Various attempts have been made to address these scalability challenges by introducing parallelism tailored to each simulator's design.  For example ns-3~\cite{riley2010ns} supports parallel simulations by by distributing networks among multiple processes communicating via MPI,~\cite{spec:mpi}, while gem5~\cite{binkert:gem5} supports concurrent execution through multiple event queues. DONS~\cite{gao2023dons} builds a clean state parallel network simulator that features Data-Oriented Design.
\sys is a generic faramwork providing building blocks to parallelize multiple existing simulators, and ensures efficient communication between each parallel component. At the same time, \sys enables end-to-end performance simulation by integrating host simulation, helps users to delve deeper into specific investigation points.
\textbf{Modular Simulator: } SimBricks~\cite{li:simbricks} introduces modular full system simulation capabilities with efficient parallelization, however, its performance is constrained by the slowest component simulator. \sys  seeks to address this bottleneck by implementing additional parallelization within each standalone simulator. pd-gem5~\cite{alian:pd-gem5} and dist-gem5~\cite{mohammad:distgem5} establish full networked systems simulation by combining multiple gem5 instances in parallel and synchronizing them through global barriers. \sys integrates various simulator types and provides a convenient orchestration framework for configuring and launching large-scale simulations. The Structual Simulation Toolkit(SST)~\cite{rodrigues:sst} presents a modular simulator enabling users to integrate components through a common interface and execute them in parallel using MPI. Unlike \sys, SST lacks the capability to profile each component, identify bottlenecks, and conduct further decomposition.

\textbf{Network Performance Estimator: }Theoretical models~\cite{zhao2023scalable,10.1145/2377677.2377747} serve as valuable tools for describing network states and estimating key metrics such as throughput and latency. While they are helpful in scenarios where network behavior can be precisely articulated through equations, they lack the packet-level visibility offered by DES. Recently researchers have turned to AI-powered estimators as an alternative to traditional theoretical models ~\cite{yang2022deepqueuenet, 10.1145/3286062.3286083,10.1145/3452296.3472926}. Unlike \sys these estimators can not provide an end-to-end application performance and requires re-learning the model for new network configurations.
\section{Conclusions}
In this paper we have introduced \sys, a system and methodology for
enabling end-to-end evaluation for large scale systems in simulation
for when physical testbeds are out of reach.
\sys uses mixed fidelity simulation to reduce resource requirements,
parallelizes component simulators through decomposition to mitigate
bottlenecks and increase simulation speed, streamlines configuring and
profiling such simulations.
In our evaluation we have shown \sys can simulate multiple hosts with
full OS and application software stacks, NICs, as part of a large
scale network of 1200 hosts, run this simulation on a single physical
machine, and complete a 20s simulation run in under four hours.
Finally, \sys drastically runs the barrier of entry for such
simulations, by enabling users to configure such simulations
comprising multiple different simulators etc. without needing
expertise for how to configure each and every simulator.
 \if \ANON 0
\fi

\bibliographystyle{plain}
\bibliography{paper,bibdb/papers,bibdb/strings,bibdb/defs}

\label{page:last}
\end{document}